\documentclass[12pt]{article}

\usepackage{scicite}

\usepackage{times}

\usepackage{graphicx}

\newenvironment{sciabstract}{%
\begin{quote} \bf}
{\end{quote}}

\title{Fish-inspired tracking of underwater turbulent plumes} 

\author
{Peter Gunnarson$^1$, John O. Dabiri$^{1,2 \ast}$\\
\\
\normalsize{$^{1}$Graduate Aerospace Laboratories, California Institute of Technology, Pasadena, CA, USA}\\
\normalsize{$^{2}$Mechanical and Civil Engineering, California Institute of Technology, Pasadena, CA, USA}\\
\\
\normalsize{$^\ast$To whom correspondence should be addressed; E-mail:  jodabiri@caltech.edu.}
}

\date{}

\begin{document} 


\baselineskip24pt


\maketitle 

\begin{sciabstract}
Autonomous ocean-exploring vehicles have begun to take advantage of onboard sensor measurements of water properties such as salinity and temperature to locate oceanic features in real time. Such targeted sampling strategies enable more rapid study of ocean environments by actively steering towards areas of high scientific value. Inspired by the ability of aquatic animals to navigate via flow sensing, this work investigates hydrodynamic cues for accomplishing targeted sampling using a palm-sized robotic swimmer. As proof-of-concept analogy for tracking hydrothermal vent plumes in the ocean, the robot is tasked with locating the center of turbulent jet flows in a 13,000-liter water tank using data from onboard pressure sensors. To learn a navigation strategy, we first implemented Reinforcement Learning (RL) on a simulated version of the robot navigating in proximity to turbulent jets. After training, the RL algorithm discovered an effective strategy for locating the jets by following transverse velocity gradients sensed by pressure sensors located on opposite sides of the robot. When implemented on the physical robot, this gradient following strategy enabled the robot to successfully locate the turbulent plumes at more than twice the rate of random searching. Additionally, we found that navigation performance improved as the distance between the pressure sensors increased, which can inform the design of distributed flow sensors in ocean robots. Our results demonstrate the effectiveness and limits of flow-based navigation for autonomously locating hydrodynamic features of interest.
\end{sciabstract}


\section*{INTRODUCTION}
The ocean is critically under-explored and under-sampled. After centuries of effort, 80\% of the seafloor remains unexplored ({\it 1\/}) and it is estimated that up to 90\% of species biodiversity in the volume of the ocean is unstudied ({\it 2\/}). Increased sampling of the ocean is vital to understanding oceanic transport processes and marine ecosystems, both of which impact global biodiversity, food supply, and climate change ({\it 3\/}). Additionally, sustained time-series observations are needed to quantify the accelerating pace of climate change due to natural variability and anthropogenic factors ({\it 4\/}).

In response to the need for increased ocean sampling, autonomous underwater vehicles (AUVs) such as undersea gliders ({\it 5\/}) and autonomous floats ({\it 6\/}) have become vital tools for in-situ sampling of ocean environments. Autonomy lowers the cost of deploying an underwater vehicle and enables exploration of larger swaths of the ocean volume by removing the need for constant communication with the surface. However, the limited range and speed of current AUVs are barriers to accomplishing widespread coverage of the ocean ({\it 7\/}).


A promising technique for increasing the effectiveness of AUVs is targeted sampling, in which robots actively seek out areas of scientific interest such as undersea thermal vents, coastal upwelling fronts, or phytoplankton patches ({\it 8\/}). Vehicles may use a variety of sensing methodologies for tracking areas of high importance. For example, an AUV used cameras and a machine vision algorithm to autonomously track animals in the midwater for hours at a time ({\it 9\/}). Another vehicle used chemical and turbidity measurements to autonomously locate an undersea thermal vent ({\it 10\/}). Salinity-sensing was used for autonomously locating and mapping the boundary of a salinity-intrusion front from the Gulf Stream ({\it 11\/}). By using onboard sensors to actively steer towards areas of high scientific value, autonomous underwater vehicles can sample information-rich locations more quickly and better allocate their limited energy supplies.

A biology-inspired approach to seeking out targets underwater is to take advantage of hydrodynamic cues present in ocean environments. For example, many aquatic animals including sea lions and catfish can hunt by sensing the wakes left behind by their prey ({\it 12, 13\/}). This ability to track animals without any visual information is advantageous for predators. If implemented in underwater vehicles, flow sensing could be a means to track oceanic features of interest when poor water and lighting conditions would otherwise obscure visual tracking. Aquatic animals use flow sensing for a variety of other tasks, including following walls ({\it 14\/}) and station keeping in the wake behind obstacles ({\it 15\/}). Each of these biological tasks could inspire similar functionalities in autonomous vehicles, e.g. autonomous navigation along the seafloor and detecting wake signatures from physical obstacles or hydrothermal vents ({\it 16\/}).

Given the potential applications of flow sensing in underwater robots, significant effort has been dedicated to the development of bio-inspired flow sensors ({\it 17\/}). For example, engineers have developed various hair-like sensors that mimic the flow-sensing mechanism of superficial neuromasts in fish lateral lines ({\it 18\/}). By measuring the deflection of micro-pillars, these sensors can detect fluid shear and infer flow velocity. The undulatory geometry of sea lion whiskers has also inspired the design of flow velocity sensors ({\it 19\/}). In addition to velocity sensing, the pressure sensing function of canal neuromasts in fish lateral lines has been mimicked using distributed pressure sensors ({\it 20\/}).

In general, these flow sensors have achieved high accuracy and sensitivity, but developing strategies to interpret these flow measurements for autonomous exploration remains an active area of research. Idealized potential flow models, often supplemented by empirical measurements or regression models, have been used for tasks such as characterizing the free stream flow ({\it 21\/}), locating dipole sources ({\it 22\/}), swimming along walls ({\it 23\/}), and vehicle state estimation ({\it 24\/}). In the context of navigation for reaching a destination or finding flow features, a variety of heuristic strategies have been studied, such as navigating using the local velocity gradient ({\it 25\/}); turning in the direction of maximum flow information ({\it 26\/}); balancing the signal of two sensors for rheotaxis ({\it 27\/}); and synchronizing swimming motion with the frequency of a Kármán Wake ({\it 28\/}). Many of these studies indicate that gradient detection with distributed flow sensors may be important for flow-based navigation (see also ({\it 29--32\/})), but unifying design principles remain unclear. 

Recently, machine learning algorithms such as Reinforcement Learning (RL) have become powerful tools for developing more complex flow-based navigation strategies. Reinforcement Learning has been used in simulated environments for locating the source of turbulent odor plumes ({\it 33\/}), following hydrodynamic trails behind simulated fish ({\it 34\/}), and for point-to-point navigation in vortical flows ({\it 35\/}). In a few cases, RL has been successfully applied to physical robots for navigation in background flow fields, such as a glider that learned to ride atmospheric thermals ({\it 31\/}) and an AUV that learned autonomous underwater target tracking ({\it 36\/}). However, the majority of RL studies are performed in silico because of the high data gathering requirements, poor interpretability, and computational complexity of neural networks ({\it 36, 37\/}). Indeed, a well-documented gap exists between successfully applying RL in simulated environments and deploying it on real systems, perhaps due to noise and factors that are difficult to model in simulations ({\it 37\/}). Trials using physical robots are needed to validate RL approaches before deployment in real-world scenarios ({\it 38\/}).

In this work, we introduce the Caltech Autonomous Reinforcement Learning robot (CARL), a palm-sized hardware and software platform for testing flow-based navigation and RL in large, controlled underwater environments. The robot is made from low-cost, off-the-shelf components and 3D-printed parts, and can swim autonomously underwater. We mounted pressure sensors at four locations around the robot to serve as flow sensors. As a proof-of-concept analogy for tracking hydrothermal vent plumes in the ocean, the robot was tasked with locating the center of turbulent jet plumes in a large water tank. A schematic overview of CARL and the tank environment is shown in Figure 1.  

To discover an effective navigation strategy, we implemented RL ({\it 39\/}) in a simulated version of the robot and tank environment (see Methods for details). Using simulated pressure measurements as inputs to a neural network, the RL algorithm successfully learned to locate the turbulent plumes. Calculating a feature importance metric ({\it 40\/}) revealed that this navigation strategy depended primarily on the lateral sensors, which provide information about the transverse velocity gradient. Using this interpretation, we transferred a simplified version of the navigation policy to the physical robot, taking into account physical sensor noise. After transferring this policy onto the physical robot, CARL located the turbulent plumes in the tank at more than double the rate of random searching, demonstrating the effectiveness of flow sensing for autonomous navigation in a physical setting. Additionally, we found that the success rate and gradient sensing ability depended greatly on the spacing of the physical sensors. An analysis of the signal-to-noise ratio suggests that the sensor spacing may limit navigation performance for physical swimmers. Our results demonstrate the effectiveness of using simulation and physical experiments in tandem to design simple but effective navigation strategies for deployment in real-world environments.

\begin{figure}
\begin{center}
\includegraphics[width=12.7cm]{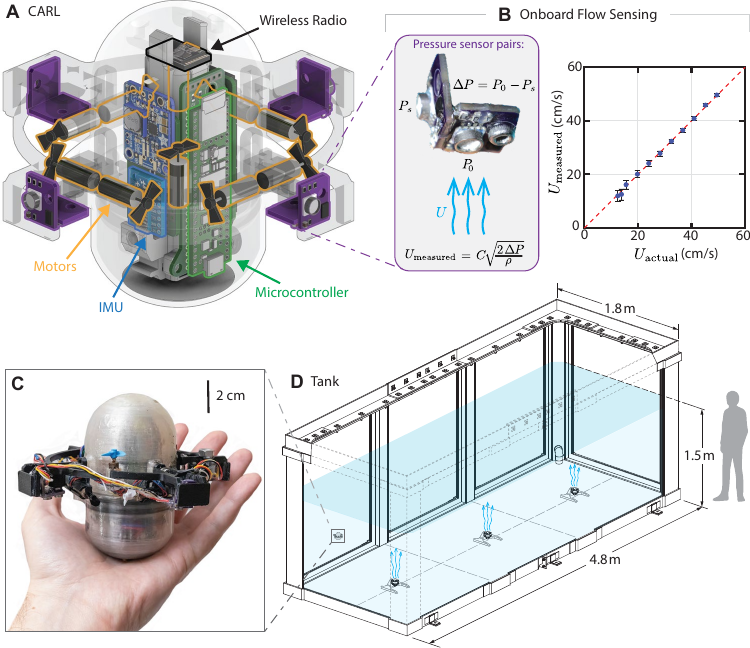}
\end{center}
\textbf{Fig 1. Schematic of the Caltech Autonomous Reinforcement Learning robot (CARL) and tank facility} \\
(\textbf{A}) Schematic of CARL, showing the arrangement of four pressure sensor pairs (see also panel \textbf{B}), an inertial measurement unit (IMU), ten motors for propulsion, and a Teensy 4.1 microcontroller for onboard processing. (\textbf{B}) Pressure sensors are arranged in pairs: a side facing pressure sensor measures the static pressure $(P_s)$, and a downward facing sensor measures the stagnation pressure $(P_0)$ from upwards flow. The difference in pressure between these sensors ($\Delta P = P_0 - P_s$) can accurately measure flow velocity, as shown in the right plot (see Methods for details). (\textbf{C}) CARL with a 2 cm scalebar. (\textbf{D}) Tank facility with a human figure for scale. Three thrusters are arranged on the bottom of the tank, which generate three vertical turbulent plumes. CARL swims throughout the tank at a fixed depth of 30 cm below the water surface.
\end{figure}

\newpage

\section*{RESULTS}

\paragraph*{Flow sensing with pressure sensors} \paragraph{}

To test flow-based navigation in a physical robot, we developed the Caltech Autonomous Reinforcement Learning robot (CARL), an autonomous underwater robotic platform. For onboard flow sensing, we mounted eight pressure sensors (MS5803-02BA, TE Connectivity) at four locations around CARL (Figure 1A). We chose this piezo-resistive micro-electromechanical system (MEMS) pressure sensor for its small size (6 mm diameter), low cost (\$16), and high precision (2.4 Pa resolution at 100 Hz). Additionally, the MS5803 sensors are manufactured with a waterproof gel coating and are already deployed in the ocean environments as depth sensors ({\it 41\/}). Pressure sensors mimicking the canal neuromasts of fish were selected due to their great mechanical robustness and commercial availability compared to micro-pillar velocity sensors that mimic superficial neuromasts. The convenience and performance of pressure sensors makes them an attractive option for experimental studies: these sensors and other piezo sensor arrays have been used in several previous works for flow characterization and robot state estimation in quiescent flow ({\it 24, 42, 43\/}).

We arranged the pressure sensors in pairs to form downward-facing pitot tubes, in which one sensor is exposed to impinging vertical flow and the other is shielded by a 3D-printed cover (see Figure 1B). In this arrangement, the difference in pressure between the exposed and shielded sensors can be used to measure the upwards flow velocity component at these four locations. We verified the accuracy of these sensors for detecting steady flow in water channel test (see Methods).

Because the center of mass of CARL was located below the center of buoyancy, the exposed pressure sensors maintained a downward-facing orientation while swimming. The pressure sensors also functioned as depth sensors; while swimming, CARL attempted to maintain a constant depth using a proportional integral derivative (PID) control loop running at 50 Hz using this depth measurement and the vertically oriented motors (see Methods).

\paragraph*{Navigation task and underwater testing environment} \paragraph{}

As a proof-of-concept analogy for tracking underwater thermal vents, which create large, turbulent jet plumes with flow velocities on the order of 1 m$\,$s$^{-1}$ ({\it 16\/}), we tasked CARL with locating the core of the turbulent jet plumes in a 1.8 m deep, 1.8 m wide, and 4.8 m long water tank. We mounted three thrusters (Blue Robotics T200) on the bottom of the tank, as shown in Figure 1D. The three thrusters were equally spaced 1.6 m apart along the centerline of the tank length, which created three distinct turbulent plumes. The thrusters have a diameter of approximately $D$ = 10 cm, which was used as a reference length scale in the subsequent analysis. To reduce the complexity of this navigation problem, CARL swam at a fixed depth of 30 cm, which is approximately $12D$ above the thrusters on the bottom of the tank. At this depth, the turbulent plumes have a spread to a diameter of approximately $5D$, which is significantly larger than the size of CARL. By swimming at a fixed depth, the navigation problem becomes effectively two-dimensional, which simplifies the possible action space for CARL. Additionally, onboard flow measurements are minimally impacted by the motion of CARL because the flow due to horizontal robot motion is perpendicular to the vertical flow-sensing orientation of the pressure sensors. The downward facing orientation of the sensors enabled CARL to detect flow from the upward-facing thrusters at the bottom of the tank and measure the mean velocity profile of the plume (see Methods for details).  

\newpage

\paragraph*{Learning a navigation policy in a simulated environment} \paragraph{}

To develop a navigation strategy for autonomously locating the turbulent jets, we first trained a navigation policy using Reinforcement Learning in a simulated environment (see Methods). By training in a virtual environment, hyperparameters such as the reward function, network size and action space could be rapidly tested and fine-tuned. For example, our simulated environment trained using 600 episodes generated over the course of several minutes, which would take several hours to accomplish with CARL in the physical tank. We used the policy learned in simulation as a starting point to design an interpretable and robust navigation policy that can function on the physical version of CARL.

In the virtual environment, we modeled CARL as a massless point swimmer that swam at a constant speed in a 2D plane to emulate swimming at a constant depth in the physical tank. Because the mean flow of the jet was normal to the swimming direction of CARL, we made the simplifying assumption that the trajectory of CARL was unaffected by the surrounding flow field and vice-versa, which eliminated the need to solve for the background flow field at each time step. The tank dimensions, sensor spacing, swimming speed, and radius of the simulated swimmer were all matched with their physical counterparts.

To train a navigation policy, we implemented the Double Deep Q-network (DDQN) Reinforcement Learning algorithm ({\it 39\/}), which seeks to predict the Q-values, i.e. the value of an action in a particular state, and selects the actions with the highest predicted Q-values (see Methods for implementation details). For the state, we used one time step of simulated pressure measurements ($\Delta P_{\mathrm{front}},\Delta P_{\mathrm{left}},\Delta P_{\mathrm{back}},$ and $\Delta P_{\mathrm{right}}$ as shown in Figure 2A). In the physical CARL robot, pressure measurements were time-averaged over 0.3 seconds to reduce sensor noise, and the navigation policy also updated at this interval. This duration of averaging was chosen to maximally reduce sensor noise without introducing an excessive delay in navigation. For example, with a swimming speed of ~20 cm$\,$s$^{-1}$, it typically took CARL 2.5 seconds to cross the width of the turbulent plume. 

To simulate these pressure measurements, we included four virtual sensors that measured the square of the vertical velocity component of a simulated turbulent jet flow field with Gaussian sensor noise scaled to approximate the noise of the physical sensors. Details of the simulated flow field are shown in Methods. To avoid simple memorization of the turbulent flow field by the neural network, the simulation starting time and the starting location for CARL was randomized at the start of each episode.

Both the virtual swimmer and CARL could swim in five possible directions as shown in Figure 2A. All actions included a component in the forward swimming direction to ensure exploration of the tank environment. After eventually running into the side walls of the tank, CARL turned around by a random angle and continued swimming. Each collision with the walls of the tank constituted the start and end of an episode. 

Initially, the swimmer selected between the five possible actions at each time step with equal probability. After training on 600 episodes of exploration with random actions, the swimmer navigated by choosing the action with the highest Q-value as predicted by the neural network. For reproducibility, we trained the policy using ten different initial random seeds. Details of the reward function and training procedure are shown in Methods.

\paragraph*{Navigation strategy learned in simulation} \paragraph{}

After training, we recorded the navigation success rate of the swimmer over 2,000 test episodes. As a baseline navigation policy for comparison, we used random exploration, in which the swimmer swam in straight lines and turned around by a random angle after colliding with the side walls of the tank. 

Given the noise of the sensors, a flow may only be detectable if the swimmer is within approximately 2.5$D$ of a jet center, which comprises only 7\% of the total area of the tank (see Methods for details). To reject episodes in which the swimmer never encountered a turbulent plume, we defined the navigation success rate as the probability of successfully finding a jet center given that a turbulent plume was encountered. For consistency with subsequent sections, we defined a plume encounter as occurring if the difference in flow measured by any two sensors was greater than twice the root-mean-square (RMS) of the sensor noise. 

The results are plotted in Figure 2E, and an example trajectory that shows the simulated swimmer steering towards the center of a turbulent plume is plotted in Figure 2D. Whereas random exploration resulted in a successful navigation rate of 14\%, the RL policy located the center of a jet in 34\% of plume encounters. Using flow sensing, the learned navigation policy significantly outperformed random searching.

\begin{figure}
\begin{center}
\includegraphics[width=\textwidth]{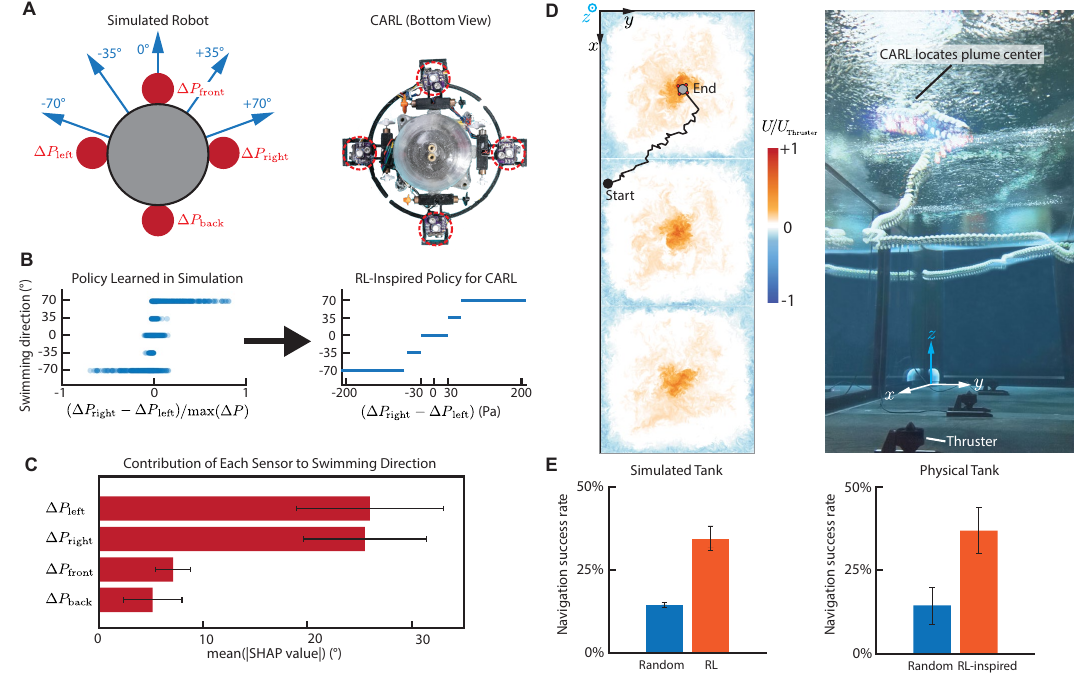}
\end{center}
\textbf{Fig 2. Navigation policy learned in simulation transfers to physical robot} \\
(\textbf{A}) A simulated version of CARL (Left) was modelled as a point swimmer with distributed sensors that emulate those on CARL (Right). (\textbf{B}) Left: navigation policy learned in simulated is plotted at each timestep, showing a clear dependence on the transverse gradient ($\Delta P_{\mathrm{right}} - \Delta P_{\mathrm{left}}$). Right: a simplified version of the learned policy captured the same behavior but was scaled to account for the physical sensor noise ($\sigma_{\Delta P} = 15$ Pa). (\textbf{C}) The left and right sensors have the largest impact on the learned swimming direction, as quantified by the SHAP values. Error bars are the standard deviation from training with 10 random seeds. (\textbf{D}) Example successful episodes in the simulated flow field (Left) and in the phsyical tank (Right). The background wall behind the tank is blurred for readability. (\textbf{E}) Left: after training, the simulated swimmer more than doubles the probability of locating a jet compared to random exploration. Error bars represent the standard deviation of the success rate after training with 10 random seeds. Right: using the RL-inspired policy, CARL achieved a similar improvement in jet-finding performance. Error bars are 95\% confidence intervals using the Wilson score interval ($N$ = 340 and 365).
\end{figure}

To investigate how the learned policy uses sensor measurements to locate the plumes, we computed SHAP (SHapley Additive exPlanations) values for each sensor. SHAP values are based on Shapley values from game theory and quantify the contribution of an input to a model to its output ({\it 40\/}). In Figure 2C, we plot the mean absolute SHAP value for each sensor, which represents the importance of each sensor averaged across all time steps of the test episodes. According to the SHAP values, the left and right sensors contributed significantly more to the learned policy than the front and back sensors.

To understand the relative importance of the left and right pressure sensor pairs, we plotted the swimming direction versus the difference of the left and right sensors ($\Delta P_{\mathrm{right}} - \Delta P_{\mathrm{left}}$) for every timestep in the 2,000 test episodes (Figure 2B). From this plot, the learned policy appears to involve turning in the direction of the sensor experiencing higher flow. Given that the flow velocity is higher in the center of turbulent plume, navigation towards faster flow leads the robot to the center of the turbulent plume. In situations where there is not a significant left-to-right velocity gradient (i.e. $\Delta P_{\mathrm{right}} - \Delta P_{\mathrm{left}}$ is close to zero), the learned policy does not appear to strongly correlate with measurements from any of the four sensors. Without a large gradient signal, the sensor inputs were dominated by turbulent fluctuations and simulated sensor noise, and therefore no action had a detectable advantage over any other. Because $\Delta P_{\mathrm{right}}$ and $\Delta P_{\mathrm{left}}$ measure vertical flow velocity at two spatially separated locations, the difference between these two sensors can be interpreted as representing a transverse velocity gradient, or a gradient in the direction perpendicular to forward swimming. Since all actions were biased towards forward swimming, a velocity gradient in the front-back direction may be less important: the swimmer will tend to explore in the forwards direction with random swimming. 

In summary, the virtual swimmer learned an effective navigation strategy for locating the turbulent jet plumes, which depends primarily on the transverse velocity gradient. However, it is not guaranteed that this policy generalizes to a physical robot with noisy sensors and encountering a real-world turbulent flow. Therefore, we next tested the learned policy in the physical tank using CARL.

\paragraph*{Plume localization with the physical CARL} \paragraph*{}

To test the gradient-based navigation strategy on the physical CARL robot, we designed an ``RL-inspired'' navigation policy, which is a simplified version of the policy learned in simulation that takes into account the sensor noise of the physical pressure sensors. A plot of the RL-inspired policy is shown in Figure 2B. In summary, if the difference between the left and right sensors is less than two standard deviations of the sensor noise, i.e. the signal-to-noise ratio (SNR) is less than two, CARL swims straight forward. Otherwise, CARL swims in the direction of larger transverse velocity gradient, as in the virtually-learned policy. The RL-inspired policy was programmed with simple if-then statements, which is computationally simple to evaluate onboard the microcontroller, particularly when compared to evaluating the output of neural networks (see Supplementary Note 1 for pseudocode). Hang et al. ({\it 34\/}) also used RL to handcraft a simple navigation policy for following hydrodynamic trails. Here, we additionally take into account limitations of a physical robot such as sensor noise and limited computation.

We conducted navigation tests in the physical tank to compare the RL-inspired policy with random navigation. Because the physical CARL lacks knowledge of its absolute position in the tank, we used depth as a proxy to determine successful location of the turbulent plume. Specifically, when CARL entered the center of the turbulent plume, the two diving motors on CARL were unable to overcome the mean flow, and CARL was pushed upwards by several centimeters. This effect only occurred in the center of the plume; in the edges of the turbulent plume the diving motors were strong enough to maintain a constant depth. The change in depth also served as a measurement of success that was independent of the flow sensors. The start and end of each episode were marked by CARL colliding with the walls of the tank, which was detected with the onboard IMU. 

An example successful trajectory is plotted in Figure 2D. Initially, while the sensed gradient was below the SNR threshold, CARL swam straight and explored the tank. When CARL detected a gradient, CARL turned in the direction of the transverse gradient in order to locate the plume above the labelled thruster. Additional example trajectories are shown in Supplementary Movies S1 and S2.

The navigation results averaged over two hours of swimming in the tank for each policy (approximately 350 episodes), are plotted in Figure 2E. Random exploration successfully located the jet center in 14\% of plume encounters, while the RL-inspired policy achieved a significantly higher success rate of 37\%. By sensing a transverse velocity gradient, CARL was able to locate turbulent jets autonomously.

\newpage

\paragraph*{Effect of sensor spacing on navigation performance} \paragraph{}

The learned navigation policy relied on detecting flow gradients using physically separated flow sensors in the presence of sensor noise and turbulent fluctuations. Therefore, the success of gradient-based navigation may be limited by the minimum detectable gradient over the background noise floor. For example, the distance between flow sensors ($L$) may be an important design consideration, since a robot with sensors spaced father apart may have a greater sensitivity to spatial gradients in the background flow but may be unable to detect flow structures smaller than $L$. 

To vary the minimum detectable gradient on CARL, we created two additional sensor mounts with reduced distance between the pressure sensors (see Figure 3A). In general, flow structures of size $L$ or smaller may be undetected or spatially aliased when sampled by two sensors. However, for this experimental setup, the mean flow profile of the turbulent plume was significantly larger than CARL for all sensor configurations (see Figure 3B). Reducing the sensor separation therefore reduced the difference in mean flow measured by the left and right sensors, negatively impacting sensitivity to flow gradients. Because the navigation policy depends on the inherent noise of the sensors which is independent of the sensor spacing or background turbulence, we tested navigation performance using the same policy for all three sensor mounts and compared the results with random exploration. For each case, we recorded the navigation performance over two hours of swimming, or approximately 350 episodes. 

In Figure 3C, we plot the plume detection rate, which we define as the chance of detecting any gradient signal above two times the noise floor of the sensors in a given episode. Since the swimming direction only changed if this threshold was exceeded, the random and RL-inspired swimming behaviors are expected to be indistinguishable. The plume detection rate increased with a greater distance between the sensors.

\begin{figure}
\begin{center}
\includegraphics[width=5in]{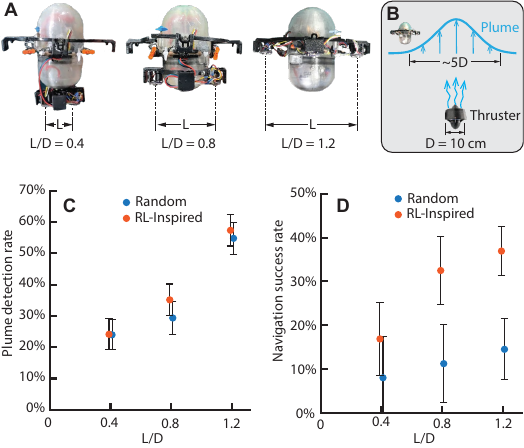}
\end{center}
\textbf{Fig 3. Navigation performance depends on sensor spacing} \\
(\textbf{A}) We created three versions of CARL with sensors of varying spacing $L$. (\textbf{B}) At the swimming depth of CARL, the mean profile of the turbulent plume is approximately 5$D$, according to flow measurements (see Methods). (\textbf{C}) The chance of CARL detecting a transverse velocity gradient above the noise floor increased with sensor separation. (\textbf{D}) The ability to locate a turbulent plume also increased with larger sensor separation. Error bars indicate 95\% confidence intervals using the Wilson score interval ($N$ for each data point ranges from 313 to 365).
\end{figure}

Additionally, we plot the navigation success rate in Figure 3D. The success rate significantly increased as the sensor spacing increases. In the case of the greatest sensor spacing ($L/D = 1.2$), the RL-inspired policy successfully located the jet cores 37\% of the time, compared with only 17\% for the smallest sensor spacing ($L/D = 0.4$). As the sensors are placed more closely together, the navigation ability became increasing similar to that of random exploration. 

The dependence of navigation success rate on sensor spacing may be explained by the signal-to-noise ratio (SNR) of the measured flow gradient. Since CARL is significantly smaller than the mean flow profile, we use a linear mean background flow (i.e. $\mathrm{d}\bar{U}/\mathrm{d}x$ is constant) to model the SNR. Because the pressure sensor pairs measured the dynamic pressure of incoming flow (i.e. Figure 1B), the gradient signal used for navigation is equal to:

\begin{equation}
\Delta P_{\mathrm{right}} - \Delta P_{\mathrm{left}} = \frac{1}{2} \rho \left(\bar{U}^2_{\mathrm{left}} - \bar{U}^2_{\mathrm{right}}\right)  = \rho \bar{U}_{\mathrm{avg}} \frac{\mathrm{d}\bar{U}}{\mathrm{d}x} L \, ,
\end{equation}

\noindent where $\bar{U}_{\mathrm{avg}}$ is the average of $\bar{U}_{\mathrm{Left}}$ and $\bar{U}_{\mathrm{Right}}$. According to this model, the signal scales with $L$: farther apart sensors experience a greater velocity differential, and therefore $\Delta P_{\mathrm{right}} - \Delta P_{\mathrm{left}}$ scales with $L$. The inherent noise of the sensors was fixed, depending only on the time-averaging window used. Noise from turbulent fluctuations may vary spatially and temporally, but for simplicity, we assumed the magnitude of turbulent fluctuations was similar for both sensors because CARL is small relative to the size of the mean flow profile. Under these assumptions, the signal is proportional to $L$ while the noise sources remain constant with sensor spacing. This may explain the increase in navigation ability as the sensor spacing increases and may also contribute to the plume detection rate.

Another potential effect is that farther apart sensors are more likely to encounter a plume during straight-line swimming. During straight swimming, the area swept out between the two sensors is proportional to $L$. Therefore, the chance of any sensor encountering a plume will scale with $L$, provided that any relevant flow features are not smaller than the distance between the sensors. Given that the turbulent plumes only occupy ~7\% of the area of the tank, an increase in swept area may partially explain the increased plume detection rate as the sensor spacing increases. However, the the navigation success rate should not depend on this effect, since it only counts episodes in which a jet is already detected.

\section*{DISCUSSION}

The effectiveness of the learned navigation policy in both simulations and physical experiments demonstrates the applicability of RL for solving flow-based navigation problems and identifying useful hydrodynamic cues. In this study, the robot learned to follow transverse flow gradients to localize the turbulent jet plumes. Applying interpretability metrics to the policy learned in simulation enabled us to simplify and adapt the learned policy for deployment in a physical robot, maintaining overall effectiveness while taking into account computational constraints and the noise of the physical sensors. The success of CARL at locating turbulent plumes demonstrates the potential for targeted sampling of real-world flow features with onboard flow sensing. 

The importance of the transverse velocity gradient suggests that onboard flow sensors may provide higher utility when arranged perpendicular to the direction of swimming. Such sensing arrangements are not uncommon in animals. For example, zebrafish were shown to require flow sensing on both sides of their body in order to detect flow gradients for avoiding walls ({\it 30\/}). Swimming in the direction of greater flow is similar to the turning strategy employed by Braitenburg vehicles ({\it 27, 44\/}) and a virtual robot that tracked the wake behind simulated fish ({\it 34\/}). While a turbulent plume is qualitatively different than the vortex shedding wake produced by an animal, both types of flows involve spreading wake-like structures and intermittent eddies. Future work could investigate tracking vortex shedding dominated wakes in a physical tank. 

The dependence of navigation performance on sensor spacing suggests that the signal-to-noise ratio is limiting for navigating via flow gradients. Therefore, the navigation strategy employed by CARL may be most effective when mean flow gradients are significant, such as in close proximity to a hydrothermal plume or the turbulent wake behind an obstacle. In addition to using more accurate flow sensors, a larger separation between sensors can improve the signal-to-noise ratio, provided that the gradients of interest are larger than the gap between sensors. Fish lateral lines often extend over the entire body, which may be advantageous for increasing sensitivity to flow gradients. Additionally, using an array of sensors to measure flow at many locations could provide additional information for navigation. In fish, distributed flow sensing can indicate flow direction and location of oscillating sources, e.g., other animals ({\it 45\/}).

Onboard distributed pressure sensing offers a convenient, low-cost, and low-power method for measuring flow gradients. If deployed in an ocean environment, calibration procedures such as those outlined in ({\it 46\/}) may be needed to compensate for water temperature and atmospheric pressure variations to achieve same flow-sensing accuracy as in lab studies. Additionally, CARL navigated using the transverse gradient, which was perpendicular to the motion of the robot. Given the three-dimensional nature of underwater navigation in ocean environments, swimming may not be limited to directions normal to all onboard sensors, which may induce flow signals during swimming. Bio-inspired robots which swim using undulatory motion also generate confounding flow signals from body motion. If sensors readings are coupled with the swimmer’s motion, pre-calculated models such as those used in ({\it 47\/}) could be implemented to disentangle pressure signals from self-motion and external stimuli. 

CARL located the turbulent plumes using only a single time step of pressure measurements. However, several studies have shown that neural network architectures with memory, such as Long Short-Term Memory (LSTM) networks, demonstrate performance improvements for tasks such as locating the source of odor plumes ({\it 33\/}) and controlling the lift of a wing in turbulent conditions ({\it 48\/}). Memory may be particularly useful in turbulent flows, which are inherently intermittent and time-varying. Turbulent fluctuations could themselves be a useful signal for navigation. For example, turbulent fluctuations have been used as a signal for distinguishing between flows ({\it 42\/}), and there is evidence that aquatic animals sense intermittency at the edge of turbulent odor plumes for locating the source of the plume ({\it 49\/}). Other flow signals, such as the static pressure or vorticity, are coupled with the flow velocity and could be useful hydrodynamic cues for navigation (e.g. ({\it 34\/})). Additionally, memory itself could be used for gradient sensing, since directionality can be encoded in a time series ({\it 50\/}). Finally, training using physically collected data or directly onboard an underwater robot may improve performance and allow for real-time adaptation to changing flow conditions.

\section*{METHODS}

\paragraph*{Components and construction of CARL} \paragraph*{}

The primary components and construction of CARL are shown in Figure 1A-C. The hull and other structural components were 3D printed using PLA (polylactic acid) and made watertight with a coating of two-part epoxy. The hull was approximately 6 cm in diameter and 10 cm tall, and was weighted on the bottom with tungsten powder mixed into epoxy resin to passively stabilize the robot in its vertical orientation. The majority of electronic components were housed inside the hull and accessible via a removable cap that formed a watertight seal with two O-rings. A Teensy 4.1 microcontroller provided onboard control and computation, and data were stored onboard using a micro-SD card. A wireless communication module (Songhe NRF24l01+ mini) was used to send and receive commands and data from CARL while on the surface. Because the water in the tank blocked wireless communication, CARL operated autonomously while underwater. Additionally, an inertial measurement unit (IMU, MPU-6050, TDK InvenSense) measured acceleration and angular rotation rate for wall impact detection and active rotational stabilization. A 12-watt-hour lithium-ion battery (Samsung 35E 18650) provided power to the robot and was replaced after an hour of swimming in typical experimental conditions. The electronic components were mounted on custom printed circuit boards (PCBs).

For propulsion, CARL was equipped with ten brushed DC motors (Crazepony 615 17500KV) with corresponding propellers. Originally intended for propulsion in air, the propellers were cut by hand to a diameter of 15 mm from the original diameter of 42 mm to account for the higher torque required for operation in water. The motors receive power from the battery through H-bridge motor drivers (Texas Instruments DRV8833) and are controlled using pulse width modulation (PWM) signals from the Teensy microcontroller. The arrangement of the motors is shown in Fig 1A. 

Eight of the motors were arranged horizontally to enable translation in both horizontal axes (front-back, left-right) and to control the rotation of the robot in the vertical axis. A 50 Hz PID control loop running on the microcontroller reads the angular rate from the onboard IMU and controls the rotation of the robot to ensure straight swimming and accurate turning. Two propellers are mounted vertically for diving. CARL is slightly positively buoyant and rises to the surface when these two propellers are turned off. This arrangement of ten motors was chosen to provide full control over translation in all three axes. Wires from all exterior electronic components were passed through the hull and sealed with epoxy.

\paragraph*{Flow sensing with pressure sensors} \paragraph*{}

We mounted pressure sensors at four locations on CARL, as shown in Figure 1A-B. The pressure sensors record an absolute pressure measurement, resulting in a signal largely dominated by the hydrostatic pressure, which varies with depth. However, taking the difference between the exposed and covered pressure sensors effectively cancels out the hydrostatic pressure. Because CARL maintains an upright orientation, the sensors maintain a constant depth relative to each other, requiring only that an initial offset is subtracted at the beginning of each episode. The exposed and covered pressure sensors at each location on CARL are mounted close together, which reduces any change relative depth due to small wobbling motions during swimming from the turbulent jets.

To validate the flow sensing capabilities of these sensors, we mounted CARL horizontally in a water channel and recorded the pressure from two sensors in steady flow conditions with speeds ranging from approximately 11 cm$\,$s$^{-1}$ to 49 cm$\,$s$^{-1}$ (Figure 4A). For this water channel test, one sensor was mounted on the side of CARL and another sensor was mounted to point into the free stream flow as shown in Figure 4B. In all other experiments, the sensors were arranged according to the Pitot-tube arrangement previously described in Figure 1. The initial offset was recorded for each sensor in zero flow conditions and subtracted from subsequent measurements. 

\begin{figure}
\begin{center}
\includegraphics[width=5in]{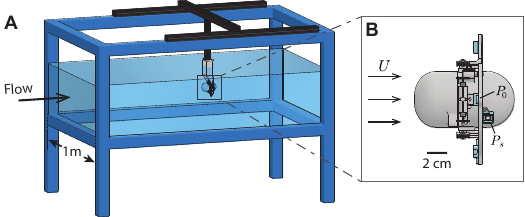}
\end{center}
\textbf{Fig 4. Validating flow sensing in a water channel} \\
(\textbf{A}) Schematic setup showing CARL placed in the water channel to validate the flow sensors. Because flow is horizontal, CARL is oriented left-to-right. (\textbf{B}) Illustration of the incoming flow and location of the flow-facing and side-facing pressure sensors, which measure $P_{0}$ and $P_{s}$, respectively.
\end{figure}

Because the exposed sensor is pointed into the freestream flow, it is expected to experience stagnation flow and a pressure increase equal to the dynamic pressure, $1/2 \rho U^2$. According to an idealized potential flow model with no body wake such as the one used in ({\it 51\/}), a side-facing sensor should experience a pressure decrease of approximately twice the dynamic pressure as the flow accelerates over the body of the robot. Therefore, the difference in pressure between the sensors, or $\Delta P$, is expected to equal $3/2 \rho U^2$, or equivalently:

\begin{equation}
U = C \sqrt{\frac{2 \Delta P} {\rho}} \, ,
\end{equation}

\noindent
where $C$ is a Pitot Tube calibration constant equal to $\sqrt{1/3}$. Using this equation, we computed the measured velocity and compared it with the flow velocity of the water channel in Figure 1B. Using a fitted constant of $C = \sqrt{0.301}$, which differs from the predicted constant by only 5\%, the sensors were able to accurately measure the freestream flow velocity, demonstrating the efficacy of these pressure sensors for quantifying flow. In all free-swimming navigation experiments, the signals from the sensors were left as pressure measurements to streamline onboard signal processing and simplify the signal-to-noise ratio analysis.

During free-swimming tests, the pressure sensors also functioned as depth sensors. The depth was estimated by taking an average of the pressure measured by the four side-facing sensors and applying the equation for the hydrostatic pressure of an incompressible fluid:

\begin{equation}
    h = P_{\mathrm{s}}/ \rho g
\end{equation}

While the flow in the tank was not static, we estimate that the hydrostatic pressure at the typical swimming depth of 30 cm was more than an order of magnitude greater than dynamic pressure created by the highest measured flow impinging on CARL. 

\paragraph*{Flow measurement of the turbulent plumes} \paragraph*{}

For comparison with the simulated jets, we measured the physical turbulent plumes using particle image velocimetry (PIV). We seeded the tank with 100-micron silver-coated hollow glass particles (AGSL150-30TRD, Potters Industries) and illuminated a cross section of the jet with a 532 nm continuous-wave laser (6 watt, Laserglow Technologies) and sheet-forming optical assembly (see Fig 5A for the experimental setup). A high speed camera (Edgertronic SC2, Sanstreak Corp) with a fixed lens (Nikon 50mm f/1.8 D) recorded the flow at 200 frames per second for 15 seconds. We processed the images in MATLAB using PIVlab ({\it 52\/}). We measured the flow at a range of throttle values from 15\% to 100\%, and plotted a snapshot of the turbulent jet flow field at 75\% throttle in Figure 5C. Flow speeds of the physical jet are on the order of 1 m$\,$s$^{-1}$, and the flow is turbulent. The thrusters were powered with a 12 V DC power adapter, and the speed was controlled via PWM outputted by a Teensy 4.1 microcontroller. During navigation tests, the thruster operated at 35\% throttle to generate a strong enough flow to be detectable by CARL without being too strong as to greatly disrupt swimming.

\begin{figure}
\begin{center}
\includegraphics[width=\textwidth]{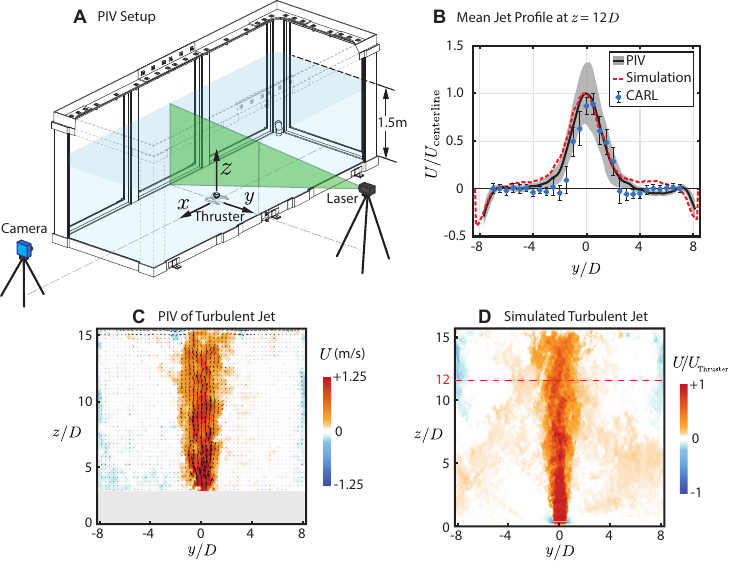}
\end{center}
\textbf{Fig 5. Measurement and simulation of the turbulent plume} \\
(\textbf{A}) Particle Image Velocimetry (PIV) setup. (\textbf{B}) Mean flow profile at a depth of 30 cm below the water surface ($z \approx 12D$) as computed from PIV, a static traverse of CARL in the tank, and the simulated jet flow. Velocity measurements from CARL are computed using equation 2 and normalized by the centerline velocity from PIV measurements ($U_{\mathrm{centerline}} \approx 0.58$ m$\,$s$^{-1}$). The error band and error bars indicate one standard deviation of the measured flow, which arises from the combination of turbulent fluctuations and measurement noise. (\textbf{C}) Snapshot of the turbulent jet PIV. Arrows are plotted to indicate flow direction and magnitude. Colors indicate the magnitude of the vertical velocity component $U$. (\textbf{D}) Snapshot of the simulated turbulent jet flow. Colors indicate the magnitude of the vertical velocity component normalized by the velocity imposed at the jet outlet ($U_\mathrm{Thruster}$). A dotted red line at $z = 12D$ indicates the depth of the mean profiles plotted in (\textbf{B}).
\end{figure}

To verify that the pressure sensors on CARL can detect the jet profile, we statically mounted CARL in the tank at a depth of 30 cm and recorded the time-averaged velocity at locations along the plume profile using equation 2. The time-averaged vertical flow speeds measured by the pressure sensors are compared with the mean flow profile measured with PIV in Figure 5B and show good agreement. This demonstrates that in a time-averaged sense, the flow sensors onboard CARL can detect and measure the mean profile of the turbulent jet flow. Error bars on the measurements from CARL indicate error due to the combination of turbulence and inherent sensor noise, which highlights a limitation to the sensing abilities of CARL for this flow field: if CARL is greater than $\sim\!2.5D$ away from the center of the plume, the noise is greater in magnitude than the mean flow. 

\paragraph*{Simulated turbulent jet flow} \paragraph*{}

To generate a flow field for the simulated robot that models the turbulent jet flow in the physical tank, we simulated the jet flow using a lattice Boltzmann solver with a Smagorinsky-Lilly subgrid turbulence model (FluidX3D software ({\it 53, 54\/})). Taking advantage of the fact that the three thrusters are equally spaced in the tank, we simulated one-third of the tank volume with one thruster. For simplicity, we applied a no-slip condition to all boundaries including the free surface. The thruster was modeled by enforcing a uniform velocity condition on a disk of magnitude $U_{\mathrm{thruster}}$ with the same dimensions as the thruster outlet and applying a jet Reynolds number of 100,000 to approximately match the flow speeds found by PIV measurements of the physical jet. The flow was simulated on a uniform 652 by 622 by 612 grid on an NVIDIA RTX 3090 GPU, resulting in a grid cell size of $\Delta x \approx D/36$. We generated 150 seconds of the turbulent jet flow, which is significantly longer than it takes CARL to swim across the tank, and used a 2D slice of the vertical velocity component at the swimming depth of CARL for training (see Figure 2D). A snapshot of the turbulent jet flow is shown in Figure 5D and the mean profile is plotted in Figure 5B, and show good agreement with PIV measurements of the physical jets.

\paragraph*{Reinforcement learning algorithm} \paragraph*{}

To train a navigation policy for the simulated swimmer, we implemented the Double DQN algorithm ({\it 39\/}), which seeks to optimize the actions of an agent to maximize a cumulative reward function. DDQN trains two sets of weights in a neural network to predict the Q-values, i.e. the value of an action in a particular state, and selects actions with higher predicted Q-values. DDQN as implemented in this study is limited to discrete outputs, however, this proved sufficient for the plume-finding task. For the Q-network, we used a two-layer multi-layer perceptron network with 64 softsign neurons per layer, which has been sufficiently expressive to solve flow-based navigation problems in previous work ({\it 35\/}). Other RL algorithms may certainly provide stability or data efficiency benefits for solving tasks involving fluids ({\it 55\/}), but because fluid-robot interactions are ignored in this simplified simulation, exploring the environment is computationally inexpensive and thus data efficiency is not critical.

The virtual CARL received a reward if the center of a jet was successfully reached, which occured if CARL swam within a diameter of 2$D$ from the center of a turbulent plume. In previous work for navigating in flow fields ({\it 35, 56\/}), intermediate rewards such as the change in distance to the target were necessary to supply a consistent reward signal during training. Untrained swimmers could not overcome the strong background flow and reach the target, and would therefore not receive a reward signal without the inclusion of an intermediate reward. Here, in the absence of strong currents that restrict swimming, random exploration occasionally resulted in successfully finding the center of a plume, therefore intermediate rewards were not required.

\section*{References}

\begin{enumerate}

\item A. C. Wölfl, H. Snaith, S. Amirebrahimi, C.W. Devey, B. Dorschel, V. Ferrini, V.A. Huvenne, M. Jakobsson, J. Jencks, G. Johnston, G. Lamarche, Seafloor mapping -- the challenge of a truly global ocean bathymetry. {\it Front. Mar. Sci.\/} {\bf 6}, 283 (2019).
\item M. L. Reaka-Kudla, Known and unknown biodiversity, risk of extinction and conservation strategy in the sea, in {\it Waters in peril} (2001), pp. 19-33. 
\item T. DeVries, C. Le Quéré, O. Andrews, S. Berthet, J. Hauck, T. Ilyina, P. Landschützer, A. Lenton, I. D. Lima, M. Nowicki, J. Schwinger, Decadal trends in the ocean carbon sink. {\it Proc. Natl. Acad. Sc.\/} {\bf 24}, 11646--11651 (2019).
\item H. M. Benway, L. Lorenzoni, A. E. White, B. Fiedler, N. .M. Levine, D. P. Nicholson, M. D. DeGrandpre, H. M. Sosik, M. J. Church, T. D. O’brien, M.Leinen, Ocean time series observations of changing marine ecosystems: an era of integration, synthesis, and societal applications. {\it Front. Mar. Sci.\/} {\bf 6}, 393 (2019).
\item D. L. Rudnick, Ocean research enabled by underwater gliders. {\it Ann. Rev. Mar. Sci.\/} {\bf 8}, 519--541 (2016).
\item A. P. S. Wong, et al., Argo data 1999--2019: Two million temperature-salinity profiles and subsurface velocity observations from a global array of profiling floats. {\it Front. Mar. Sci.\/} {\bf 7}, 700 (2020).
\item E. Zereik, M. Bibuli, N. Mišković, P. Ridao, A. Pascoal, Challenges and future trends in marine robotics. {\it Annu. Rev. Control.\/} {\bf 46}, 350--368 (2018).
\item Y. Zhang, J. P. Ryan, B. Kieft, B. W. Hobson, R. S. McEwen, M. A. Godin, J. B. Harvey, B. Barone, J. G. Bellingham, J. M. Birch, C. A. Scholin, F. P. Chavez, Targeted sampling by autonomous underwater vehicles. {\it Front. Mar. Sci.\/} {\bf 6}, 415 (2019).
\item K. Katija, P. L. D. Roberts, J. Daniels, A. Lapides, K. Barnard, M. Risi, B. Y. Ranaan, B. G. Woodward, J. Takahashi, Visual tracking of deepwater animals using machine learning-controlled robotic underwater vehicles, in {\it Proceedings of the IEEE/CVF Winter Conference on Applications of Computer Vision} (2021), pp. 860-869.
\item V. Preston, G. Flaspohler, A. P. M. Michel, J. W. Fisher III, N. Roy, Robotic planning under uncertainty in spatiotemporal environments in expeditionary science. {\it arXiv preprint arXiv:2206.01364\/} (2022).
\item Y. Zhang, N. Yoder, B. Kieft, A. Kukulya, B. W. Hobson, S. Ryan, Glen G. Gawarkiewicz, Autonomous tracking of salinity-intrusion fronts by a long-range autonomous underwater vehicle. {\it IEEE J. Ocean Eng.\/} {\bf 4}, 950--958 (2022).
\item K. Pohlmann, F. W. Grasso, T. Breithaupt, Tracking wakes: The nocturnal predatory strategy of piscivorous catfish. {\it Proc. Natl. Acad. Sc.\/} {\bf 98}, 7371--7374 (2001).
\item G. Dehnhardt, B. Mauck, W. Hanke, H. Bleckmann, Hydrodynamic trail-following in harbor seals (Phoca vitulina). {\it Science.\/} {\bf 293}, 102--104 (2001).
\item P. Patton, S. Windsor, S. Coombs, Active wall following by Mexican blind cavefish (Astyanax mexicanus). {\it J. Comp. Physiol. A.\/} {\bf 196}, 853--867 (2010).
\item J. C. Liao, A review of fish swimming mechanics and behaviour in altered flows. {\it Philos. Trans. R. Soc. B.\/} {\bf 362}, 1973--1993 (2007).
\item L. N. Germanovich, R. S. Hurt, J. E. Smith, G. Genc, R. P. Lowell, Measuring fluid flow and heat output in seafloor hydrothermal environments. {\it J. Geophys. Res. Solid Earth.\/} {\bf 120}, 8031--8055 (2015).
\item H. Ko, G. Lauder, R. Nagpal, The role of hydrodynamics in collective motions of fish schools and bioinspired underwater robots. {\it J. R. Soc.\/} {\bf 20}, 20230357 (2023).
\item M. Bora, A. G. P. Kottapalli, J. Miao, M. S. Triantafyllou, Sensing the flow beneath the fins. {\it Bioinspir. Biomim.\/} {\bf 13}, 025002 (2018).
\item X. Zheng, A. M. Kamat, M. Cao, A. G. P. Kottapalli, Creating underwater vision through wavy whiskers: a review of the flow-sensing mechanisms and biomimetic potential of seal whiskers. {\it J. R. Soc.\/} {\bf 18}, 20210629 (2021).
\item Y. Zhai, X. Zheng, G. Xie, Fish lateral line inspired flow sensors and flow-aided control: a review. {\it J. Bionic Eng.\/} {\bf 18}, 264--291 (2021).
\item L. DeVries, F. D. Lagor, H. Lei, X. Tan, D. A. Paley, Distributed flow estimation and closed-loop control of an underwater vehicle with a multi-modal artificial lateral line. {\it Bioinspir. Biomim.\/} {\bf 10}, 025002 (2015).
\item A. Dagamseh, R. Wiegerink, T. Lammerink, G. Krijnen, Imaging dipole flow sources using an artificial lateral-line system made of biomimetic hair flow sensors. {\it J. R. Soc.\/} {\bf 10}, 20130162 (2013).
\item W.-K. Yen, J. Guo, Wall following control of a robotic fish using dynamic pressure, in {\it OCEANS} (2016), Shanghai, pp. 1-7. IEEE.
\item X. Zheng, W. Wang, M. Xiong, G. Xie, Online state estimation of a fin-actuated underwater robot using artificial lateral line system. {\it IEEE Transactions on Robotics.\/} {\bf 36}, 472--487 (2020).
\item R. Monthiller, A. Loisy, M. A. Koehl, B. Favier, C. Eloy, Surfing on turbulence: A strategy for planktonic navigation. {\it Phys. Rev. Lett.\/} {\bf 129}, 64502 (2022).
\item B. Colvert, G. Liu, H. Dong, E. Kanso, Flowtaxis in the wakes of oscillating airfoils. {\it Theor. Comput. Fluid Dyn.\/} {\bf 34}, 545--556 (2020).
\item T. Salumäe, I. Rañó, O. Akanyeti, M. Kruusmaa, Against the flow: A Braitenberg controller for a fish robot, in {\it 2012 IEEE International Conference on Robotics and Automation} (2012), pp. 4210-4215. IEEE.
\item J. Ježov, O. Akanyeti, L. D. Chambers, M. Kruusmaa, Sensing oscillations in unsteady flow for better robotic swimming efficiency, in {\it 2012 IEEE International Conference on Systems, Man, and Cybernetics} (2012), pp. 91-96. IEEE.
\item B. Colvert, K. Chen, E. Kanso, Local flow characterization using bioinspired sensory information. {\it J. Fluid Mech.\/} {\bf 818}, 366--381 (2017).
\item P. Oteiza, I. Odstrcil, G. Lauder, R. Portugues, F. Engert, A novel mechanism for mechanosensory-based rheotaxis in larval zebrafish. {\it Nature.\/} {\bf 547}, 445--448 (2017).
\item G. Reddy, J. Wong-Ng, A. Celani, T. J. Sejnowski, M. Vergassola, Glider soaring via reinforcement learning in the field. {\it Nature.\/} {\bf 562}, 236--239 (2018).
\item P. Weber, Optimal flow sensing for schooling swimmers. {\it Biomimetics.\/} {\bf 5}, 10 (2020).
\item S. H. Singh, F. van Breugel, R. P. N. Rao, B. W. Brunton, Emergent behaviour and neural dynamics in artificial agents tracking odour plumes. {\it Nat. Mach. Intell.\/} {\bf 5}, 58--70 (2023).
\item H. Hang, Y. Jiao, S. Heydari, F. Ling, J. Merel, E. Kanso, Interpretable and generalizable strategies for stably following hydrodynamic trails. {\it bioRxiv\/}, pp. 2023-12 (2023).
\item P. Gunnarson, I. Mandralis, G. Novati, P. Koumoutsakos, J. O. Dabiri, Learning efficient navigation in vortical flow fields. {\it Nat. Commun.\/} {\bf 12}, 7143 (2021).
\item I. Masmitja, M. Martin, T. O’Reilly, B. Kieft, N. Palomeras, J. Navarro, K. Katija, Dynamic robotic tracking of underwater targets using reinforcement learning. {\it Science Robotics.\/} {\bf 8}, eade7811 (2023).
\item G. Dulac-Arnold, N. Levine, D. J. Mankowitz, J. Li, C. Paduraru, S. Gowal, T. Hester, Challenges of real-world reinforcement learning: definitions, benchmarks and analysis. {\it Mach. Learn.\/} {\bf 110}, 2419--2468 (2021).
\item K. Hasselmann, A. Ligot, J. Ruddick, M. Birattari, Empirical assessment and comparison of neuro-evolutionary methods for the automatic off-line design of robot swarms. {\it Nat. Commun.\/} {\bf 12}, 4345 (2021).
\item H. van Hasselt, A. Guez, D. Silver,, Deep reinforcement learning with double Q-learning, in {\it Proceedings of the AAAI Conference on Artificial Intelligence} (2016) Vol. 30, No. 1.
\item S. M. Lundberg, S. -I. Lee, A unified approach to interpreting model predictions, in {\it Advances in Neural Information Processing Systems} (2017), 30.
\item P. A. Beddows, E. K. Mallon, Cave pearl data logger: A flexible arduino-based logging platform for long-term monitoring in harsh environments. {\it Sensors.\/} {\bf 18}, 530 (2018).
\item R. Venturelli, O. Akanyeti, F. Visentin, J.  Ježov, L. D. Chambers, G. Toming, J. Brown, M. Kruusmaa, W. M. Megill, P. Fiorini, Hydrodynamic pressure sensing with an artificial lateral line in steady and unsteady flows. {\it Bioinspir. Biomim.\/} {\bf 7}, 036004 (2012).
\item M. Asadnia, A. G. P. Kottapalli, Z. Shen, J. Miao, M. Triantafyllou, Flexible and surface-mountable piezoelectric sensor arrays for underwater sensing in marine vehicles. {\it IEEE Sensors J.\/} {\bf 13}, 3918--3925 (2013).
\item V. Braitenberg, {\it Vehicles: Experiments in Synthetic Psychology} (MIT press, Cambridge, MA, 1986)
\item H. Bleckmann, R. Zelick, Lateral line system of fish. {\it Integrative Zoology.\/} {\bf 4}, 13--25 (2009).
\item N. Strokina, J. -K. Kämäräinen, J. A. Tuhtan, J. F. Fuentes-Pérez and M. Kruusmaa, Joint estimation of bulk flow velocity and angle using a lateral line probe. {\it IEEE Trans. Instrum. Meas.\/} {\bf 65}, 601--613 (2016).
\item O. Akanyeti, L. D. Chambers, J. Ježov, J. Brown, R. Venturelli, M. Kruusmaa, W. M. Megill, P. Fiorini, Self-motion effects on hydrodynamic pressure sensing: part I. Forward–backward motion. {\it Bioinspir. Biomim.\/} {\bf 8}, 026001 (2013).
\item P. I. Renn, M. Gharib, Machine learning for flow-informed aerodynamic control in turbulent wind conditions. {\it Commun. Eng.\/} {\bf 1}, 1--9 (2022).
\item R. T. Michaelis, K. W. Leathers, Y. V. Bobkov, B. W. Ache, J. C. Principe, R. Baharloo, I. M. Park, M. A. Reidenbach, Odor tracking in aquatic organisms: The importance of temporal and spatial intermittency of the turbulent plume. {\it Scientific Reports.\/} {\bf 10}, 7961 (2020).
\item N. Kadakia, M. Demi, B. T. Michaelis, B. D. DeAngelix, M. A. Reidenbach, D. A. Clarck, T. Emonet, Odour motion sensing enhances navigation of complex plumes. {\it Nature.\/} {\bf 611}, 754--761 (2022).
\item X. Zheng, W. Wang, L. Li, G. Xie, Artificial lateral line based relative state estimation between an upstream oscillating fin and a downstream robotic fish. {\it Bioinspir. Biomim.\/} {\bf 16}, 016012 (2020).
\item W. Thielicke, R. Sonntag, Particle image velocimetry for MATLAB: Accuracy and enhanced algorithms in PIVlab. {\it J. Open Res. Softw.\/} {\bf 9}, 12 (2021).
\item M. Lehmann, thesis, {\it Computational study of microplastic transport at the water-air interface with a memory-optimized lattice Boltzmann method}, University of Bayreuth, Bavaria, Germany (2023).
\item M. Lehmann, Esoteric pull and esoteric push: Two simple in-place streaming schemes for the lattice boltzmann method on GPUs. {\it Computation.\/} {\bf 10}, 92 (2022).
\item S. Berger, A. A. Ramo, V. Guillet, T. Lahire, B. Martin, T. Jardin, E. Rachelson, M. Bauerheim, Reliability assessment of off-policy deep reinforcement learning: A benchmark for aerodynamics. {\it Data-Centric Engineering.\/} {\bf 5}, e2 (2024).
\item L. Biferale, F. Bonaccorso, M. Buzzicotti, P. Clark Di Leoni, K. Gustavsson, Zermelo’s problem: Optimal point-to-point navigation in 2D turbulent flows using reinforcement learning. {\it Chaos.\/} {\bf 29}, 29 (2019).

\end{enumerate}

\newpage

\section*{Acknowledgments}
\textbf{Funding:} This work was supported by the National Science Foundation Alan T. Waterman Award and NSF Graduate Research Fellowship Grant No. DGE 1745301. \textbf{Author contributions:}  P.G. and J.O.D. conceived of project, P.G. conducted experiments, P.G. and J.O.D. analyzed results and wrote paper. \textbf{Competing interests:} The authors declare that they have no competing financial interests. \textbf{Data and materials availability:} The data collected by CARL, code for simulating the turbulent jets, and RL algorithm used in this study are available at https://doi.org/10.22002/da2jr-yx734.

\section*{Supplementary materials}
Movie S1. Time-lapse video of CARL swimming in the tank using the RL-inspired policy. Video is sped up at 40 times speed. \\
\noindent
Movie S2. Example trajectories from four successful episodes in which CARL used the RL-inspired policy to locate turbulent plumes. \\
Supplementary Note 1. Pseudocode of the RL-inspired navigation policy.

\clearpage

\end{document}